%% file: plckesz_followup.tex
\documentclass[numberedappendix]{emulateapj} 
\pdfoutput=1
\usepackage{amsmath,natbib,graphicx}
\usepackage{epsf}
\usepackage{epstopdf}
\usepackage{epsfig}
\usepackage{rotating}
\usepackage{booktabs}

\DeclareGraphicsExtensions{.jpg,.pdf,.png,.eps,.ps}
\graphicspath{{FIGURES/}}

\newcommand{\tabd}{\ensuremath{\dagger}}

\newcommand{\plcktwenty}{PLCKESZ G225.92-19.99}
\newcommand{\plckthirty}{PLCKESZ G264.41+19.48}
\newcommand{\plckseventyone}{PLCKESZ G283.16-22.93}
\newcommand{\plckseventyfive}{PLCKESZ G304.84-41.42}
\newcommand{\plckwhopper}{PLCKESZ G255.62-46.16}
\newcommand{\planck}{{\it Planck}}
\newcommand{\ROSAT}{{\it ROSAT}}

\newcommand{\sptwhopper}{SPT-CL J0411-4819}

\newcommand{\sntwenty}{13.8}
\newcommand{\snthirty}{6.3}
\newcommand{\snseventyone}{9.9}
\newcommand{\snseventyfive}{12.2}
\newcommand{\snwhopper}{14.8}

\newcommand{\ztwenty}{0.46}
\newcommand{\zthirty}{0.24}
\newcommand{\zseventyone}{0.45}
\newcommand{\zseventyfive}{0.41}
\newcommand{\zwhopper}{0.42}

\def\KICPChicago{1}
\def\PhysicsUChicago{2}
\def\UChicago{3}
\def\Munich{4}
\def\MIT{5}
\def\NCSA{6}
\def\ExcellenceCluster{7}
\def\EFIChicago{8}
\def\UAlabama{9}
\def\CfA{10}
\def\AAUChicago{11}
\def\Argonne{12}
\def\PUC{13}
\def\McGill{14}
\def\Illinois{15}
\def\Berkeley{16}
\def\UFlorida{17}
\def\Colorado{18}
\def\NASA{19}
\def\Davis{20}
\def\LBNL{21}
\def\UArizona{22}
\def\Michigan{23}
\def\MPE{24}
\def\CaseWestern{25}
\def\Caltech{26}
\def\Harvard{27}
\def\STScI{28}
\def\SAIC{29}
\def\Yale{30}

\begin{document}                                                                                
                                                                                                
\title{South Pole Telescope Detections of The Previously Unconfirmed \textit{PLANCK} Early SZ Clusters in the Southern Hemisphere}

\author{
K.~Story\altaffilmark{\KICPChicago,\PhysicsUChicago},
K.~A.~Aird\altaffilmark{\UChicago},
K.~Andersson\altaffilmark{\Munich,\MIT},
R.~Armstrong\altaffilmark{\NCSA},
G.~Bazin\altaffilmark{\Munich,\ExcellenceCluster},
B.~A.~Benson\altaffilmark{\KICPChicago,\EFIChicago},
L.~E.~Bleem\altaffilmark{\KICPChicago,\PhysicsUChicago},
M. Bonamente\altaffilmark{\UAlabama},
M.~Brodwin\altaffilmark{\CfA},
J.~E.~Carlstrom\altaffilmark{\KICPChicago,\PhysicsUChicago,\EFIChicago,\AAUChicago,\Argonne}, 
C.~L.~Chang\altaffilmark{\KICPChicago,\EFIChicago,\Argonne},
A.~Clocchiatti\altaffilmark{\PUC},
T.~M.~Crawford\altaffilmark{\KICPChicago,\AAUChicago},
A.~T.~Crites\altaffilmark{\KICPChicago,\AAUChicago},
T.~de~Haan\altaffilmark{\McGill},
S.~Desai\altaffilmark{\NCSA,\Illinois},
M.~A.~Dobbs\altaffilmark{\McGill},
J.~P.~Dudley\altaffilmark{\McGill},
R.~J.~Foley\altaffilmark{\CfA}, 
E.~M.~George\altaffilmark{\Berkeley},
M.~D.~Gladders\altaffilmark{\KICPChicago,\AAUChicago},
A.~H.~Gonzalez\altaffilmark{\UFlorida},
N.~W.~Halverson\altaffilmark{\Colorado},
F.~W.~High\altaffilmark{\KICPChicago,\AAUChicago}, 
G.~P.~Holder\altaffilmark{\McGill},
W.~L.~Holzapfel\altaffilmark{\Berkeley},
S.~Hoover\altaffilmark{\KICPChicago,\EFIChicago},
J.~D.~Hrubes\altaffilmark{\UChicago},
M.~Joy\altaffilmark{\NASA},
R.~Keisler\altaffilmark{\KICPChicago,\PhysicsUChicago},
L.~Knox\altaffilmark{\Davis},
A.~T.~Lee\altaffilmark{\Berkeley,\LBNL},
E.~M.~Leitch\altaffilmark{\KICPChicago,\AAUChicago},
M.~Lueker\altaffilmark{\Berkeley},
D.~Luong-Van\altaffilmark{\UChicago},
D.~P.~Marrone\altaffilmark{\UArizona},
J.~J.~McMahon\altaffilmark{\KICPChicago,\EFIChicago,\Michigan},
J.~Mehl\altaffilmark{\KICPChicago,\AAUChicago},
S.~S.~Meyer\altaffilmark{\KICPChicago,\PhysicsUChicago,\EFIChicago,\AAUChicago},
J.~J.~Mohr\altaffilmark{\Munich,\ExcellenceCluster,\MPE},
T.~E.~Montroy\altaffilmark{\CaseWestern},
S.~Padin\altaffilmark{\KICPChicago,\AAUChicago,\Caltech},
T.~Plagge\altaffilmark{\KICPChicago,\AAUChicago},
C.~Pryke\altaffilmark{\KICPChicago,\AAUChicago,\EFIChicago}, 
C.~L.~Reichardt\altaffilmark{\Berkeley},
A.~Rest\altaffilmark{\Harvard,\STScI},
J.~Ruel\altaffilmark{\Harvard},
J.~E.~Ruhl\altaffilmark{\CaseWestern}, 
B.~R.~Saliwanchik\altaffilmark{\CaseWestern}, 
A.~Saro\altaffilmark{\Munich},
K.~K.~Schaffer\altaffilmark{\KICPChicago,\EFIChicago,\SAIC}, 
L.~Shaw\altaffilmark{\Yale},
E.~Shirokoff\altaffilmark{\Berkeley}, 
J.~Song\altaffilmark{\Michigan},
H.~G.~Spieler\altaffilmark{\LBNL},
B.~Stalder\altaffilmark{\CfA},
Z.~Staniszewski\altaffilmark{\CaseWestern},
A.~A.~Stark\altaffilmark{\CfA}, 
C.~W.~Stubbs\altaffilmark{\Harvard,\CfA}, 
K.~Vanderlinde\altaffilmark{\McGill},
J.~D.~Vieira\altaffilmark{\KICPChicago,\PhysicsUChicago,\Caltech},
R.~Williamson\altaffilmark{\KICPChicago,\AAUChicago}, 
and
A.~Zenteno\altaffilmark{\Munich,\ExcellenceCluster}
}

\altaffiltext{\KICPChicago}{Kavli Institute for Cosmological Physics, University of Chicago, 5640 South Ellis Avenue, Chicago, IL 60637}
\altaffiltext{\PhysicsUChicago}{Department of Physics, University of Chicago, 5640 South Ellis Avenue, Chicago, IL 60637}
\altaffiltext{\UChicago}{University of Chicago, 5640 South Ellis Avenue, Chicago, IL 60637}
\altaffiltext{\Munich}{Department of Physics, Ludwig-Maximilians-Universit\"{a}t, Scheinerstr.\ 1, 81679 M\"{u}nchen, Germany}
\altaffiltext{\MIT}{MIT Kavli Institute for Astrophysics and Space Research, Massachusetts Institute of Technology, 77 Massachusetts Avenue, Cambridge, MA 02139}
\altaffiltext{\NCSA}{National Center for Supercomputing Applications, University of Illinois, 1205 West Clark Street, Urbana, IL 61801}
\altaffiltext{\ExcellenceCluster}{Excellence Cluster Universe, Boltzmannstr.\ 2, 85748 Garching, Germany}
\altaffiltext{\EFIChicago}{Enrico Fermi Institute, University of Chicago, 5640 South Ellis Avenue, Chicago, IL 60637}
\altaffiltext{\UAlabama}{Department of Physics, University of Alabama, Huntsville, AL 35812}
\altaffiltext{\CfA}{Harvard-Smithsonian Center for Astrophysics, 60 Garden Street, Cambridge, MA 02138}
\altaffiltext{\AAUChicago}{Department of Astronomy and Astrophysics, University of Chicago, 5640 South Ellis Avenue, Chicago, IL 60637}
\altaffiltext{\Argonne}{Argonne National Laboratory, HEP Division, Argonne, IL 60439}
\altaffiltext{\PUC}{Departamento de Astronomia y Astrofisica, Pontificia Universidad Catolica de Chile, Casilla 306, Santiago 22, Chile}
\altaffiltext{\McGill}{Department of Physics, McGill University, 3600 Rue University, Montreal, Quebec H3A 2T8, Canada}
\altaffiltext{\Illinois}{Department of Astronomy, University of Illinois, 1002 West Green Street, Urbana, IL 61801}
\altaffiltext{\Berkeley}{Department of Physics, University of California, Berkeley, CA 94720}
\altaffiltext{\UFlorida}{Department of Astronomy, University of Florida, Gainesville, FL 32611}
\altaffiltext{\Colorado}{Department of Astrophysical and Planetary Sciences and Department of Physics, University of Colorado, Boulder, CO 80309}
\altaffiltext{\NASA}{Department of Space Science, VP62, NASA Marshall Space Flight Center, Huntsville, AL 35812}
\altaffiltext{\Davis}{Department of Physics, University of California, One Shields Avenue, Davis, CA 95616}
\altaffiltext{\LBNL}{Physics Division, Lawrence Berkeley National Laboratory, Berkeley, CA 94720}
\altaffiltext{\UArizona}{Steward Observatory, University of Arizona, 933 North Cherry Avenue, Tucson, AZ 85721, USA}
\altaffiltext{\Michigan}{Department of Physics, University of Michigan, 450 Church Street, Ann Arbor, MI, 48109}
\altaffiltext{\MPE}{Max-Planck-Institut f\"{u}r extraterrestrische Physik, Giessenbachstr.\ 85748 Garching, Germany}
\altaffiltext{\CaseWestern}{Physics Department and CERCA, Case Western Reserve University, 10900 Euclid Ave., Cleveland, OH 44106}
\altaffiltext{\Caltech}{California Institute of Technology, 1200 E. California Blvd., Pasadena, CA 91125}
\altaffiltext{\Harvard}{Department of Physics, Harvard University, 17 Oxford Street, Cambridge, MA 02138}
\altaffiltext{\STScI}{Space Telescope Science Institute, 3700 San Martin
Dr., Baltimore, MD 21218}
\altaffiltext{\SAIC}{Liberal Arts Department, 
School of the Art Institute of Chicago, 
112 S Michigan Ave, Chicago, IL 60603}
\altaffiltext{\Yale}{Department of Physics, Yale University, P.O. Box 208210, New Haven, CT 06520-8120}

\shortauthors{Story, et al.}
\email{kstory@uchicago.edu}

\begin{abstract}
We present South Pole Telescope (SPT) observations of the five galaxy cluster candidates in the southern hemisphere which were reported as unconfirmed in the \planck\ Early Sunyaev-Zel'dovich (ESZ) sample.
One cluster candidate, \plckwhopper, is located in the 2500 deg$^2$ SPT SZ survey region and was reported previously as \sptwhopper. 
For the remaining four candidates, which are located outside of the SPT SZ survey region, we performed short, dedicated SPT observations.
Each of these four candidates was strongly detected in maps made from these observations, 
with signal-to-noise ratios ranging from \snthirty\ to \sntwenty.
We have observed these four candidates on the Magellan-Baade telescope 
and used these data to estimate cluster redshifts from the red sequence.
Resulting redshifts range from \zthirty\ to \ztwenty.
We report measurements of $Y_{0\farcm75}$, the integrated Comptonization within a $0\farcm75$ radius, for all five candidates.
We also report X-ray luminosities calculated from \ROSAT\ All-Sky Survey catalog counts,
as well as optical and improved SZ coordinates for each candidate.
The combination of SPT SZ measurements, optical red-sequence measurements, and X-ray luminosity estimates demonstrates that these five \planck\ ESZ cluster candidates do indeed correspond to real galaxy clusters with redshifts and observable properties consistent with the rest of the ESZ sample.
\end{abstract}

\maketitle


\section{Introduction}
\setcounter{footnote}{0}
In recent years, a number of new experiments have successfully discovered and characterized galaxy clusters using the thermal Sunyaev-Zel'dovich (SZ) effect \citep{sunyaev72,birkinshaw99,carlstrom02}.
The thermal SZ effect is a spectral distortion in the Cosmic Microwave Background (CMB) caused by photons inverse-Compton scattering off electrons in the hot, ionized intra-cluster medium.
The SZ surface brightness is largely independent of redshift which implies that SZ surveys with sufficient angular resolution can deliver nearly mass-limited cluster samples.

Deep SZ surveys covering over 1000 square degrees at arcminute resolution are being carried out by the South Pole Telescope \citep[SPT; ][]{carlstrom11} and the Atacama Cosmology Telescope \citep[ACT; ][]{fowler07},
and have already provided cosmologically interesting SZ-selected cluster catalogs 
(\citealt[hereafter V10]{vanderlinde10}; \citealt{marriage11}; \citealt[hereafter W11]{williamson11}).
The \planck\ collaboration has recently released an all-sky SZ-selected cluster catalog \citep[][hereafter Planck11]{planck11-5.1a_arxiv}.  
The all-sky coverage of the \planck\ survey makes it highly complementary to 
the deep, higher-resolution surveys.

In compiling their Early SZ (ESZ) catalog, the \planck\ collaboration has ensured high purity through a detection significance cut of signal-to-noise ratio (S/N) $\geq$ 6,
additional internal validation procedures, and follow-up observations of a subset of their candidates \citep{planck11-5.1b_arxiv}.
The ESZ catalog consists of 189 SZ-selected galaxy clusters, 20 of which are new detections.
Of these 20 new detections, eight were reported as unconfirmed candidates in the ESZ catalog.

Five of these unconfirmed candidates are located in the southern hemisphere and are thus observable with the SPT.
One cluster, \plckwhopper, lies in the SPT SZ survey region and was reported in W11 as \sptwhopper.  The remaining four cluster candidates lie outside the SPT SZ survey region and required dedicated observations.
By observing small patches of sky centered on the positions reported 
in Planck11, we have detected all four remaining candidates with S/N ranging from \snthirty\ to \sntwenty.
Optical observations of these four clusters were taken with the
Magellan-Baade telescope in Chile to obtain red-sequence redshifts,
and X-ray luminosities were calculated from \ROSAT\ All-Sky Survey (RASS) catalog counts.
In addition to confirming all five cluster candidates, 
we report measurements of $Y_{0\farcm75}$, the integrated Comptonization within a $0\farcm75$ radius,
red-sequence redshifts, and optical and improved SZ coordinates for the clusters.

\section{SZ Observations and Data Processing}

\subsection{SZ Observations}
\label{sec:sz_obs}
The 10-meter SPT is a millimeter/submillimeter telescope located at the NSF Amundsen-Scott South Pole Station and is optimized for observations of the CMB.
The SPT currently has a 960-pixel, three-band (95, 150, and 220 GHz) camera.
With the one-arcminute resolution and high instantaneous sensitivity of the SPT, 
any cluster at $z \gtrsim 0.2$ which was detected significantly by the \planck\ telescope 
is expected to result in a strong SPT detection after a few hours of integration.
Between January 15 and 26, 2011, we used the SPT to perform pointed observations of the four \planck\ ESZ cluster candidates in the
southern hemisphere that were not already in the SPT SZ survey region. 
These observations were similar to those described in \citet{plagge10} and involved mapping a small ($\sim$ 4 deg$^2$), square patch of sky centered on each candidate.

These patches of sky were observed with a standard SPT observing strategy
in which the telescope scans a patch back and forth in azimuth, followed by a small step in elevation,
repeating until the patch is covered.
Each patch was re-observed until a significant detection (or a high-significance non-detection) of the cluster candidate could be made.
The clusters discovered by \planck\ have typical redshifts of $z \sim 0.4$ and masses of 
$M_{500}\sim 5-10 \times 10^{14} M_{\odot}$ (cf. Fig. 18 and 19 of Planck11). 
Such a cluster will result in a $\gtrsim 10 \sigma$ SPT detection at the full SPT survey depth of $18\, \mu$K-arcmin (CMB temperature units) at 150 GHz (cf. W11), 
and this depth was our rough goal for non-detections. 
In fact, all four cluster candidates were detected long before this depth was reached, 
with a total of four to ten hours of observing time on each patch.
The patches containing \plcktwenty, \plckthirty, \plckseventyone, and \plckseventyfive\ 
were observed to field depths of 39, 29, 35, and $28\, \mu$K-arcmin at 150 GHz, respectively.

\subsection{Data Processing}
\label{sec:processing}
The data are processed through a pipeline similar to that used in W11, which we briefly describe here.
The time-ordered data (TOD) are calibrated to CMB temperature units based on observations of a galactic HII region (RCW38), as in \citet[hereafter S09]{staniszewski09}
(calibration uncertainties add a systematic uncertainty of $\approx 10\%$ to measurements of $Y_{0\farcm75}$).
The TOD are then bandpass-filtered, and correlated noise between detectors is removed.
We implement a high-pass filter by removing a fifth order Legendre polynomial and a set of Fourier modes.
We remove correlated atmospheric noise by subtracting the mean and slope of the TOD across all detectors on each of the six geometrically compact detector wedges.

The pointing for each detector is reconstructed, and the calibrated,
filtered TOD are binned into $0\farcm25$ square pixels on a Sanson-Flamsteed flat sky projection.  
The astrometry of each map is corrected by matching bright point sources which appear both in our maps and in the AT20G \citep{murphy10} and SUMSS \citep{mauch03} catalogs.

\subsection{Detecting Clusters}
\label{sec:detecting_cl}
We search for clusters in our maps by applying the same cluster extraction procedure as was used on the full SPT SZ survey fields in W11.
In this search, we do not apply any priors to the cluster positions based on the \planck\ candidate positions.
Our cluster extraction procedure is based on the application of a simultaneous spatial-spectral matched filter \citep{melin06}.
We briefly describe our inputs to this procedure here and direct the reader to W11 for full details.

Because we can predict the spectral signature of the thermal SZ effect (for a cluster of a given temperature) and the expected spatial profiles of clusters, we can combine maps from our three bands to maximize sensitivity to cluster signals and minimize noise and astrophysical contaminants.
We use an estimate of the thermal SZ spectrum in each of our three observing bands that includes a relativistic correction appropriate for an 8 keV cluster \citep{nozawa00} and uses an SZ-weighted band center.\footnote{The value of $Y_{0\farcm75}$ we would derive had we used the spectrum for a 4 keV or 10 keV cluster 
would differ by approximately $2 \%$ from the 8 keV value.}

We estimate the noise component of the matched filters independently in each frequency band in each of the four targeted observations, using the jackknife procedure described in S09.
Because these noise levels vary significantly from patch to patch, independent matched filters were constructed for each of the four targeted patches.

We model the astrophysical components of the matched filters as a combination of primary and lensed CMB fluctuations, point sources below the SPT detection threshold (bright sources are masked prior to filtering), and thermal SZ flux from clusters below the SPT detection threshold.
The power-spectrum shapes and amplitudes for these components are identical to those used in W11.

As in W11, the source template in the matched filters is described by a cored power-law SZ profile,
$\Delta T = \Delta T_0 (1+\theta^2/\theta_c^2)^{-1}$, where the normalization $\Delta T_0$ and the core radius $\theta_c$ are free parameters.
Twelve different matched filters were constructed and applied to each patch, each with a different core radius, spaced evenly between $0\farcm25$  and $3\farcm0$.
We report the maximum S/N across all filter scales for each cluster.

The detections of \plcktwenty, \plckthirty, \plckseventyone, and \plckseventyfive\ 
have S/N of \sntwenty, \snthirty, \snseventyone, and \snseventyfive, respectively.
Note that due to the highly variable map depths, these S/N measurements should not be used as a proxy for SZ flux (nor mass).
\plckwhopper\ was previously detected with S/N of \snwhopper\ in the full SPT SZ survey using this same method.

\section{Robustness of SZ Confirmations}
\label{sec:robustness}

Here we describe tests demonstrating that the SPT confirmations of \planck\ ESZ clusters are robust. 
We focus on the least significant detection in the SPT maps (\plckthirty); limits on 
the likelihood of spurious confirmation are more stringent for the other clusters.

Estimating the exact probability of a spurious SPT confirmation of a \planck\ cluster 
is complicated by our limited knowledge of the complex \planck\ selection function.  
Without an accurate estimate of the \planck\ spatial-spectral filter 
it is impossible to accurately estimate the degree of 
correlation between the astrophysical contaminants in the synthesized SZ maps 
used in the \planck\ cluster search and the corresponding SPT maps.
As the frequency coverage and angular resolution of \planck\ and SPT are 
substantially different, we proceed under the assumption of uncorrelated astrophysical contaminants.

We simulate 4000 deg$^2$ of microwave sky consisting of the primary CMB and point sources. 
We make simulated sky maps in each SPT observing band by convolving the sky signal 
with the SPT beams, applying any spatial filters that are applied to the real data 
(see Section \ref{sec:processing}),
and adding a noise realization based on the noise estimate from 
the SPT maps of the \plckthirty\ region. We then apply the same matched filters used 
to find clusters in the real SPT maps, and we record ``detections'' in
any of the filtered maps.
Zero false detections are found at S/N$>$\snthirty\ in the 4000 deg$^2$. 
Extrapolating the measured false detection rate at lower significance to S/N=\snthirty\ results in an expectation value of 0.04 in 4000 deg$^2$. 
Since we require proximity to \planck\ candidates, the effective area is much smaller---only $\sim 0.02$ deg$^2$. 
Neglecting the possibility of astrophysical confusion being correlated between the Planck and SPT observations, the chance of a false detection in any of the five confirmations is negligible at $2\times 10^{-7}$.

\section{Measuring Cluster Properties}

\subsection{Cluster Position and Integrated Comptonization}
\label{sec:cl_properties}

We estimate the position and integrated Comptonization for each
cluster with the Rapid Gridded Likelihood Evaluation (RGLE) method
described in \cite{montroy11} (in preparation). Because the 220~GHz data adds little
statistical significance, we use only 95 and 150~GHz maps.

In the RGLE, we define the likelihood of a set of parameters
$\mathcal{H}$ given our set of observed maps $D_{\nu}(\bar{x})$ as
\footnotesize
\begin{equation}
\log(P(D|\mathcal{H})) = -\frac{1}{2}\sum_{\bar{k},\nu1,\nu2}\frac{(\widetilde{D}_{\nu_1}(\bar{k})-\widetilde{s}^{\mathcal{H}}_{\nu_1}(\bar{k}))(\widetilde{D}_{\nu_2}(\bar{k})-\widetilde{s}^{\mathcal{H}}_{\nu_2}(\bar{k}))^*}{N_{\nu_1
    \nu_2}(\bar{k})},
\end{equation}
\normalsize 
where $\widetilde{D}_{\nu}(\bar{k})$ is the Fourier transform of the
map for frequency $\nu$, $\widetilde{s}^{\mathcal{H}}_{\nu}$ is the
frequency dependent Fourier transform of the cluster model for
parameters set $\mathcal{H}$, and $N_{\nu_1\nu_2}(\bar{k})$ is the
frequency dependent covariance matrix of the set of maps. We model
clusters with a cored power-law SZ profile that is exponentially cut
off for $\theta > 5 \theta_c$.  $\mathcal{H}$ contains four
parameters: the cluster center $(x,y)$, the core radius ($\theta_c$)
and the central temperature decrement $\Delta T_0$ (taken at 150~GHz).
The covariance matrix accounts for the same noise and astrophysical
components used in the matched filter analysis.

For each cluster the likelihood of a particular $\mathcal{H}$ is
evaluated on a grid in the 4 dimensional parameter space
($x,~y,~\theta_c,~\Delta T_0$).  The spacing between grid points is
$0\farcm0625$ in $x$, $y$, and $\theta_c$, and $10\mu K$ in $\Delta
T_0$. Cluster positions cover a $2\farcm5 \times 2\farcm5$ patch
centered on the matched filter position. $\Delta T_0$ is computed
between $-3000\mu K$ and $1000\mu K$, while $\theta_c$ is computed
between $3\farcs75$ and 6\farcm0.

Cluster positions are calculated by marginalizing the 4-dimensional
likelihood over the other parameters.  These positions lie within
0\farcm18 (less than the pixel size) of the best-fit matched filter
positions.

For each cluster we report $Y_{0\farcm75}$, the integrated
Comptonization of the cluster model within a fixed $0\farcm75$-radius
aperture, marginalized over $y_0$, $\theta_c$, and cluster positions.
We expect measurements of $Y_{0\farcm75}$ to be robust despite the
well known degeneracy between $\theta_c$ and the central Compton
parameter $y_0$ because the radius of integration ($0\farcm75$)
closely approximates the mean size of the 95 and 150 GHz SPT beams.
Simulations of clusters with S/N in the range of those reported here
confirm that $Y_{0\farcm75}$ is well constrained.  The choice of
cluster profiles does not add significant bias; by searching for
simulated Arnaud profile clusters \citep{arnaud09} in simulated maps
using the cored power-law SZ profile matched filter, we find the
recovered samples have a bias of less than 5\%.

\subsection{Redshifts}
\label{sec:opt_obs}
Cluster redshifts were obtained with imaging at optical/near-infrared
wavelengths following the procedure outlined in detail by
\citet{high10}, which we now summarize.  We observed the four cluster
candidates outside the SPT SZ survey region using the IMACS camera
\citep{dressler06} on the Magellan-Baade telescope at Las Campanas
Observatory, Chile, on January 27, 2011.  We operated in imaging mode
at a focal ratio of $f/2.4$.  For each cluster, we exposed for $100$
seconds through each of the $gri$ filters.  We achieved seeing of
$1\farcs3$--$1\farcs6$ in the images of \plckseventyfive, and seeing
of $0\farcs8$--$1\farcs0$ in the other three.  Images were bias-subtracted,
flat-fielded, defringed where appropriate, illumination-corrected, and
astrometrically calibrated and deprojected.  Objects were then
detected and associated between passbands with a $1\arcsec$ search
radius.  We applied the Stellar Locus Regression method of photometric
calibration \citep{high09} to the resulting catalog, including
color-term corrections.  We then searched for overdensities of red
cluster galaxies using standard red-sequence techniques.  Redshifts
were assigned using a red-sequence model that we calibrated using a
set of clusters with spectroscopic redshifts, and the overall
statistical plus systematic uncertainty is estimated to be $\sigma_z /
(1+z) \approx 2$--$3\%$.  W11 used the same procedure to obtain the
redshift of \plckwhopper, but with $VRI$ imaging data from the Swope
telescope at Las Campanas, Chile, in which we achieved seeing of
$1\farcs5$--$1\farcs6$. The redshift of \plckwhopper\ has an
uncertainty of $\sigma_z / (1+z) \approx 5\%$.

\subsection{X-ray Luminosities}
\label{sec:x-ray}
As noted in Planck11, each cluster has a nearby
X-ray association from either the \ROSAT\ All-Sky Survey Bright Source
Catalog \citep{voges99} or the \ROSAT\ All-Sky Survey Faint Source
Catalog\footnote{http://heasarc.gsfc.nasa.gov/W3Browse/rosat/rassfsc.html}.
To determine the flux and luminosity of each X-ray source, we used
XSPEC\footnote{http://heasarc.nasa.gov/xanadu/xspec/xspec11/index.html} to
model the instrument response within Position Sensitive Proportional Counter channels 52-201.  
We assumed a cluster temperature of 8 keV and a metal abundance of 0.3 times solar, 
and adjusted the normalization of the XSPEC model to match the count rates reported in the Bright Source and Faint Source catalogs.  
Towards each cluster, we assume the galactic absorbing hydrogen column density measured by the Leiden-Argentine-Bonn survey \citep{kalberla05}.  
The XSPEC ``lumin'' function was applied to the normalized spectral model to calculate the luminosity in the cluster rest frame 0.5-2.0 keV energy band, 
which we have chosen because of its relative insensitivity to the assumed X-ray temperature.  
For example, when assuming a range of gas temperatures between 4 to 12 keV, the inferred X-ray luminosity changes by $<$8\%, at a level negligible relative to the total uncertainty.
Luminosity uncertainties are computed as the quadrature sum of the red sequence redshift uncertainties and the count rate uncertainties from RASS.

\section{Results}
\label{sec:results}

\input{main_tab}

We detect a highly significant SZ signal at the expected location and spatial scale in each of the four observed patches.
No other significant SZ signals were detected in these maps.
As demonstrated in section \ref{sec:robustness}, the chance of a false detection in any of these candidates is negligible.
In addition to the SZ signal, we see clear optical counterparts for each candidate, 
and the calculated X-ray luminosities for the candidates are consistent with typical values for the \planck\ ESZ sample.
From the combination of SZ detections, optical counterparts and consistent X-ray luminosities we conclude that the SPT confirmations (and the original \planck\ detections) of all five clusters are robust.

\begin{figure*}
\centering
\includegraphics[scale=.97]{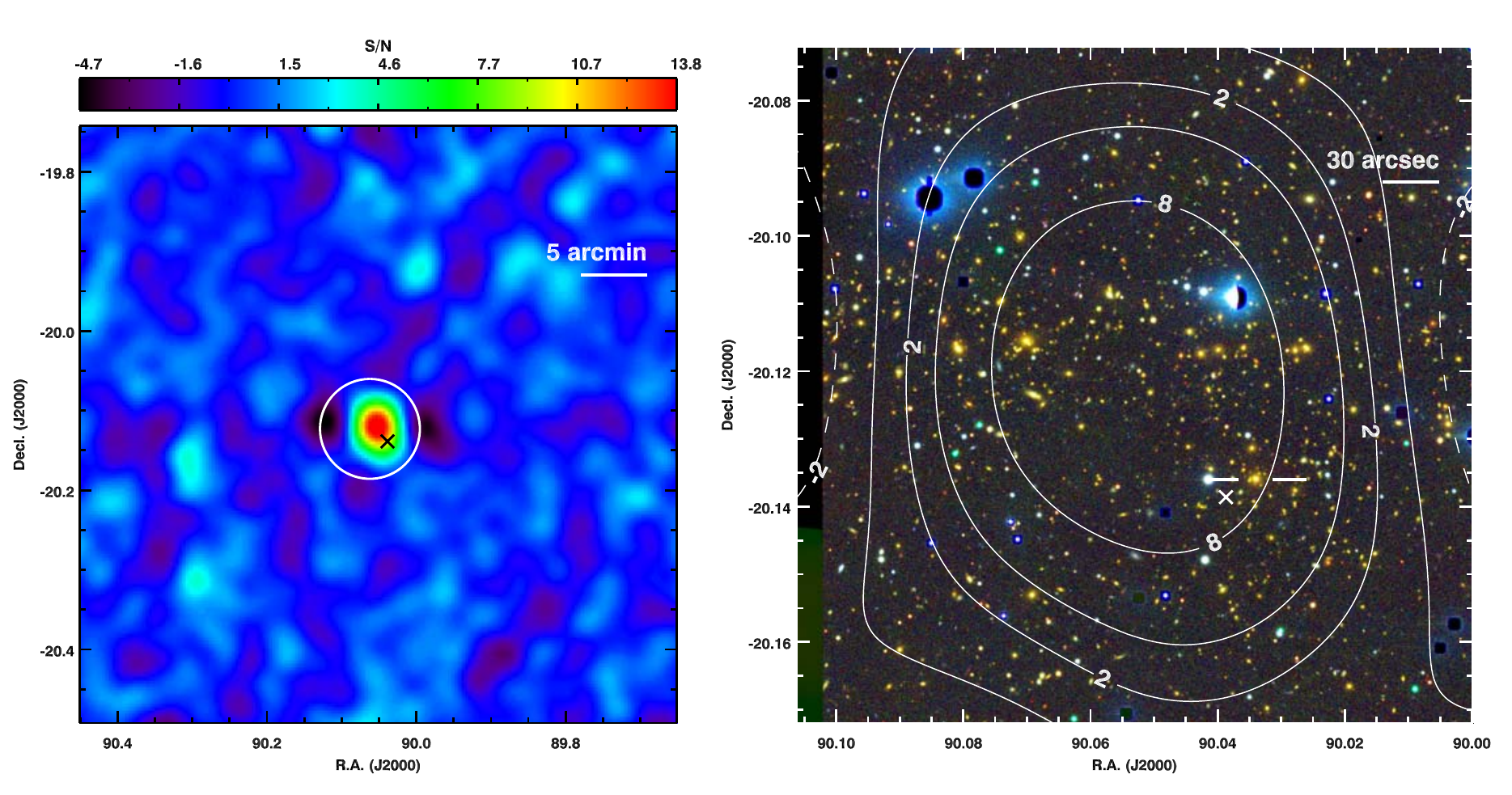}
  \caption[]{\plcktwenty\ at $z=\ztwenty$.
The SPT filtered S/N map of a 0.75 deg $\times$ 0.75 deg patch of sky centered on the cluster coordinates is shown in the left panel.
This map is the output of the spatial-spectral matched filter at the filter scale that maximized detection significance for the cluster.
The negative wings to the left and right of the cluster are the result of high-pass filtering the data.
A FWHM 7.5 arcmin beam (approximately the mean \planck\ HFI beam size) is shown for comparison, centered on the \planck\ location.
The location of the RASS X-ray association for each cluster is marked with an ``X''.
Magellan/IMACS \textit{gri} images are shown in the
optical/near-infrared panel to the right, 
where the RASS coordinate is marked with an ``X'', the BCG is
indicated by the 2 horizontal white lines, and the overlaid contours correspond to
the SPT S/N map in the left panel.}
\label{fig:pl225}
\end{figure*}

\begin{figure*}
\centering
\includegraphics[scale=.97]{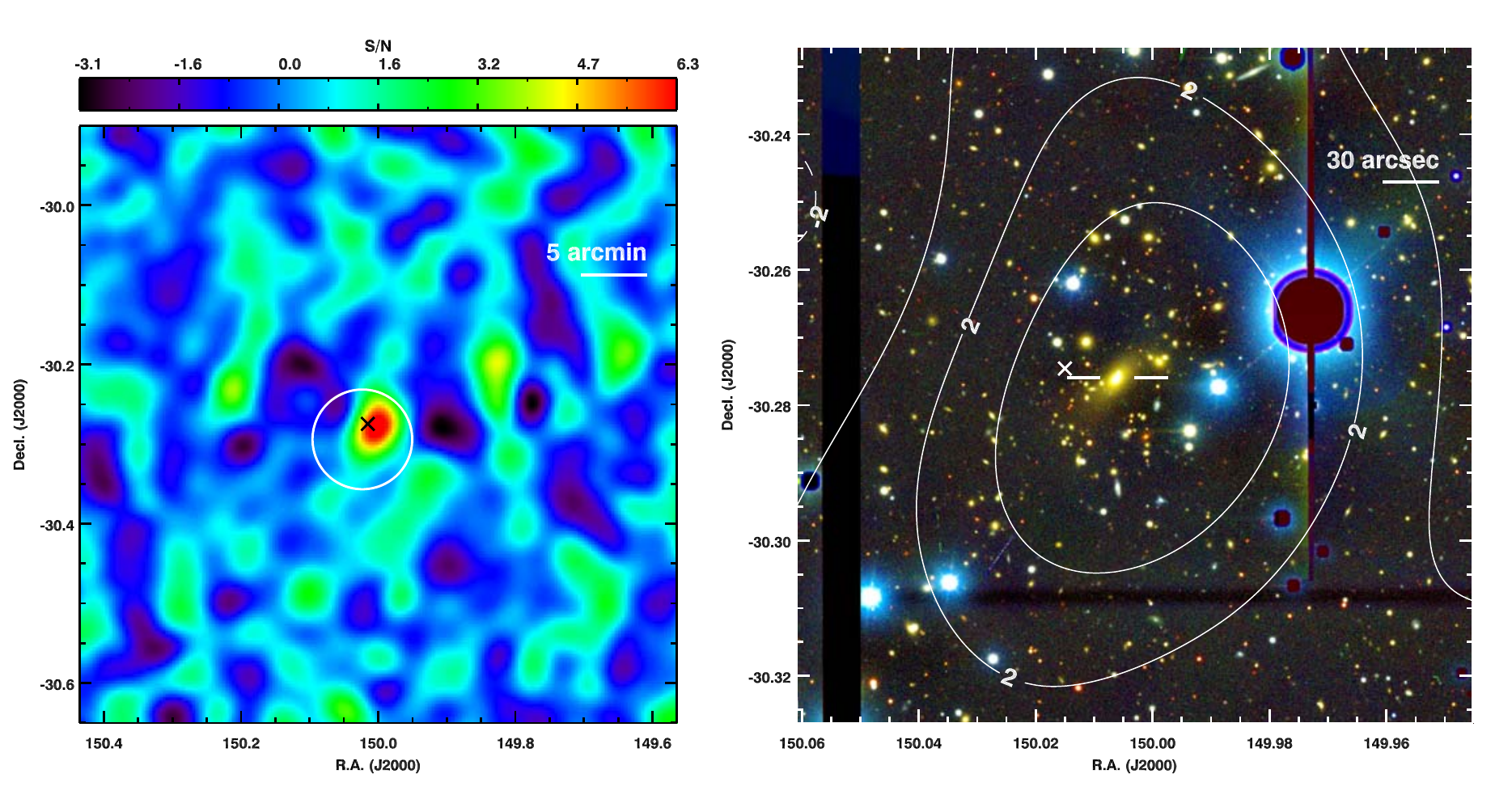}
  \caption[]{\plckthirty\ at\ $z=\zthirty$.
  See Figure \ref{fig:pl225} caption for details.
}
\label{fig:pl264}
\end{figure*}

\begin{figure*}
\centering
\includegraphics[scale=.97]{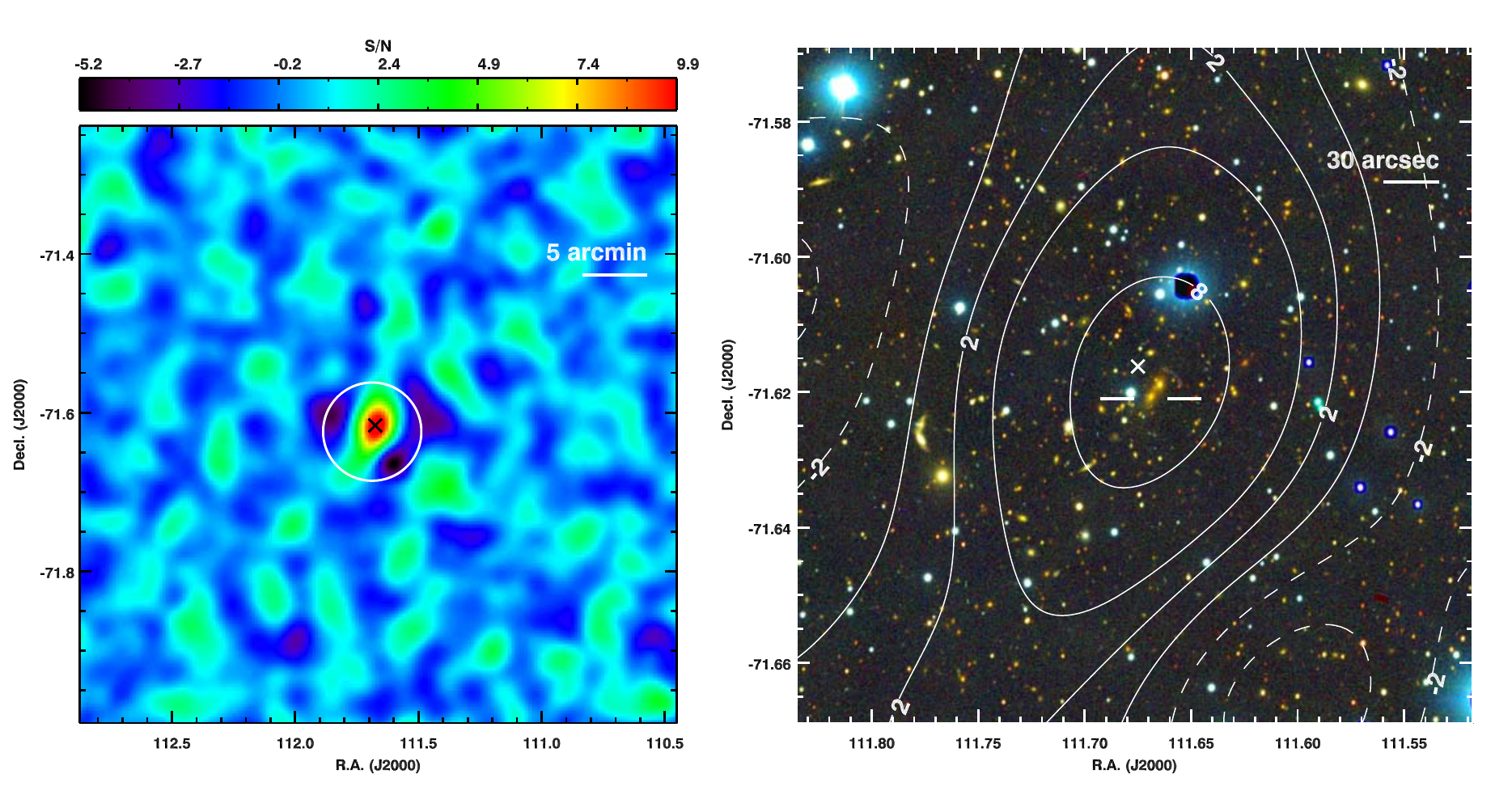}
  \caption[]{\plckseventyone\ at $z=\zseventyone$.
  See Figure \ref{fig:pl225} caption for details.
\textit{Note:} A giant, blue strong-lens arc is apparent, though faint, immediately to the northwest of the double BCG system in the optical image.  
The bright yellow extended object with a curved shape to the east of the cluster center is likely a merging galaxy system in the foreground.
}
\label{fig:pl283}
\end{figure*}

\begin{figure*}
\centering
\includegraphics[scale=.97]{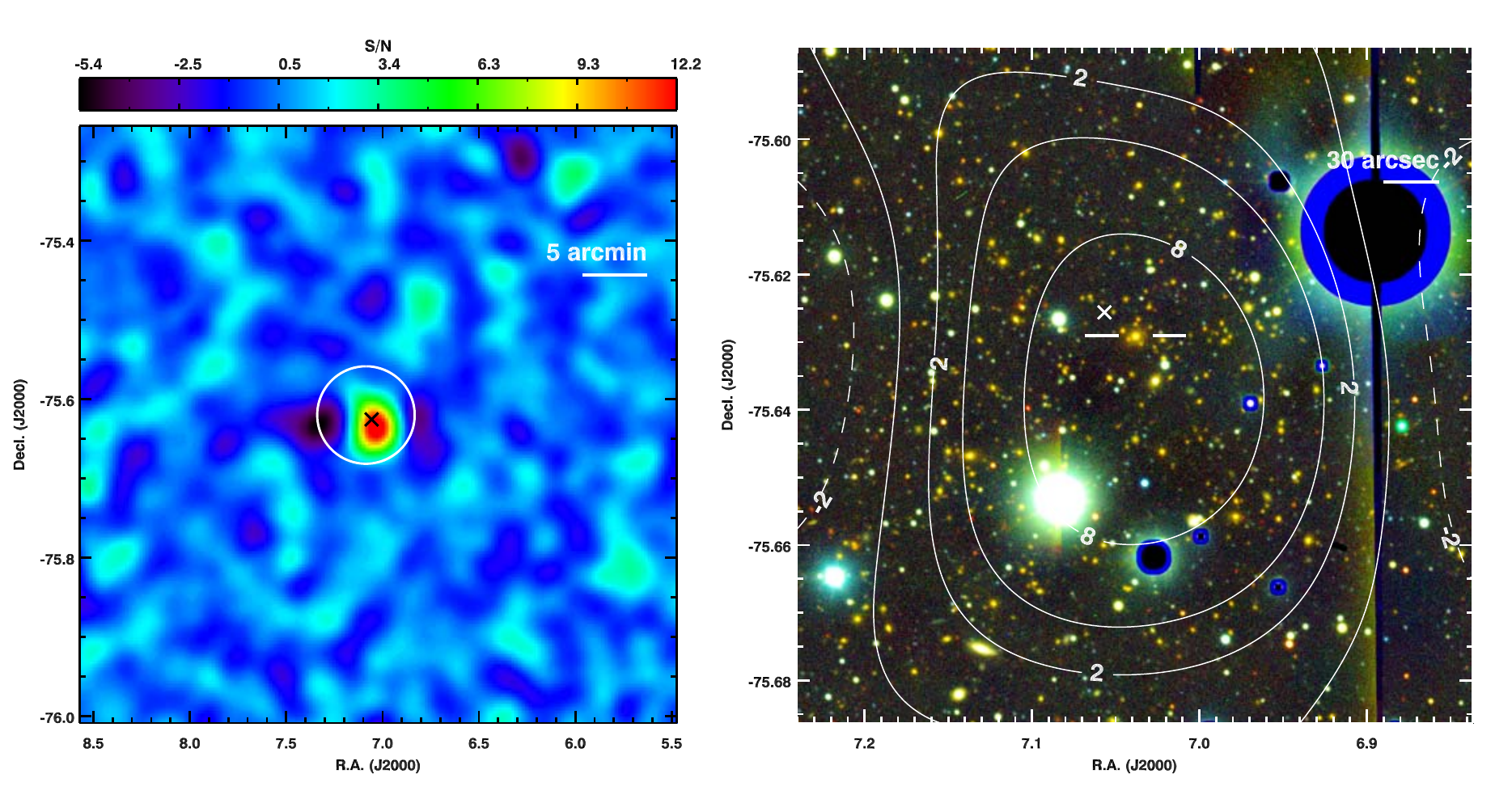}
  \caption[]{\plckseventyfive\ at $z=\zseventyfive$.
  See Figure \ref{fig:pl225} caption for details.
}
\label{fig:pl304}
\end{figure*}

For each of the four newly confirmed clusters, we show S/N maps, filtered at the preferred scale,
and optical images in Figures \ref{fig:pl225} to \ref{fig:pl304}.
From the optical image of \plcktwenty, it appears that
this cluster has two distinct components whose red-sequence redshifts are
consistent with one another, possibly indicating merger activity.

For all five clusters, we report SZ coordinates, integrated Comptonization $Y_{0\farcm75}$, red-sequence redshifts, 
X-ray luminosities, and optical coordinates of the brightest galaxy in each cluster (BCG), in Table \ref{tab:clusters}.
The BCG is selected as the brightest galaxy whose color is consistent with the cluster red-sequence within $\sim 1^\prime$ of the SPT coordinate.
Our measured SZ coordinates are consistent with the \planck\ coordinates to within the positional accuracy reported in Planck11.
The distributions of $z$ and $L_X$ are consistent with those of the new \planck\ discoveries in the ESZ catalog 
which were confirmed by dedicated XMM-Newton observations \citep{planck11-5.1b_arxiv} (cf.\ Fig. 18 and 21 of Planck11). 
From these measurements we find that these five clusters have observable properties which are consistent with the rest of the ESZ sample.


\section{Conclusions}
\label{sec:conclusion}
The recently released \planck\ ESZ sample contains five cluster candidates in the southern hemisphere which were reported as unconfirmed. 
Of these five, \plckwhopper\ was previously reported as \sptwhopper\ in W11. 
The four remaining clusters were detected at high significance in short, dedicated observations with the SPT. 
From simulations we calculate that a false detection in follow-up observations is exceedingly unlikely for any of these clusters.

From SPT SZ data we report measurements of $Y_{0\farcm75}$, the integrated Comptonization within a fixed $0\farcm75$-radius, for each cluster.
We use optical/near-infrared images of each cluster, obtained using the Magellan-Baade telescope,
to confirm the presence of overdensities of similarly colored galaxies at the position of the SZ signal and to estimate red-sequence redshifts.  
Using these redshifts and archival RASS X-ray data, we calculate the X-ray luminosity for each cluster and find that these luminosities are typical of other clusters in the \planck\ ESZ sample at these redshifts.
We additionally report optical and improved SZ coordinates for each cluster.

The combination of SPT SZ measurements, optical red-sequence measurements, and X-ray luminosity estimates demonstrates that these five \planck\ ESZ cluster candidates do indeed correspond to real galaxy clusters with redshifts and observable properties consistent with the rest of the ESZ sample. 
These measurements are consistent with the expected high purity of the \planck\ ESZ sample and
confirm all five previously unconfirmed clusters from the \planck\ ESZ sample which are observable with the SPT.

\acknowledgements

{\it Facilities:}
\facility{South Pole Telescope},
\facility{Magellan: Baade (IMACS)}

The SPT is supported by NSF grant ANT-0638937.
Partial support is also provided by NSF grant PHY-0114422, 
the Kavli Foundation, and the Gordon and Betty Moore Foundation.
The McGill group acknowledges funding from the NSERC of Canada, the Quebec Fonds de recherche sur la nature et les technologies, and the CIAR.
Galaxy cluster research is supported at Harvard by NSF grant AST-1009012 and at SAO 
by NSF grants AST-1009649 and MRI-0723073.
B. Benson acknowledges support from a KICP Fellowship,
M. Brodwin from the W. M. Keck Foundation,
A. Clocchiatti from CONICYT grants Basal CATA PFB 06/09, FONDAP No. 15010003,
R.~J. Foley from a Clay Fellowship, 
R. Keisler from NASA Hubble Fellowship grant HF-51275.01,
and
B. Stalder from the Brinson Foundation.

\bibliography{../../BIBTEX/spt.bib}

\clearpage

\end{document}

%% file: main_tab.tex
\begin{deluxetable*}{l rrcc cccr}
\tablecaption{SPT Confirmation of Planck ESZ Cluster Candidates\label{tab:clusters}}
\tablehead{
\multicolumn{1}{c}{}  & 
\multicolumn{3}{c}{SPT}	    & 
\multicolumn{3}{c}{Optical} & 
\multicolumn{1}{c}{X-ray} \\
\cmidrule(lr){2-4} \cmidrule(lr){5-7} \cmidrule(l){8-8}
\colhead{Object Name}	& 
\colhead{R.A.\tablenotemark{a}}		& 
\colhead{Decl.\tablenotemark{a}}		& 
\colhead{$Y_{0\farcm75}$\tablenotemark{b}} & 
\colhead{R.A.\tablenotemark{c}}		& 
\colhead{Decl.\tablenotemark{c}}		& 
\colhead{$z$\tablenotemark{d}}		& 
\colhead{$L_X$\tablenotemark{e}} \\
&
\colhead{[deg]} & 
\colhead{[deg]} & 
\colhead{[arcmin$^2$]}&
\colhead{[deg]} & 
\colhead{[deg]} & 
& 
\colhead{[$10^{44}$ergs/s]}
}
\startdata
\plcktwenty                       &  90.052 $\pm$ 0.002 & -20.119 $\pm$ 0.002 & 4.02 $\pm$ 0.29 &  90.0340 & -20.1358 & \ztwenty       & 5.2 $\pm$ 1.4 \\
\plckwhopper\tablenotemark{\tabd} &  62.814 $\pm$ 0.002 & -48.319 $\pm$ 0.002 & 1.89 $\pm$ 0.13 &  62.7957 & -48.3276 & \zwhopper      & 6.0 $\pm$ 1.8 \\
\plckthirty                       & 150.002 $\pm$ 0.004 & -30.276 $\pm$ 0.004 & 1.25 $\pm$ 0.23 & 150.0062 & -30.2761 & \zthirty       & 1.5 $\pm$ 0.5 \\
\plckseventyone                   & 111.672 $\pm$ 0.005 & -71.619 $\pm$ 0.003 & 2.19 $\pm$ 0.20 & 111.6687 & -71.6207 & \zseventyone   & 2.9 $\pm$ 0.9 \\
\plckseventyfive                  &   7.034 $\pm$ 0.006 & -75.635 $\pm$ 0.002 & 2.25 $\pm$ 0.17 &   7.0379 & -75.6292 & \zseventyfive  & 3.8 $\pm$ 1.2
\enddata
\tablenotetext{\tabd}{\ Cluster \plckwhopper\ was previously reported as \sptwhopper\ in W11.}
\tablenotetext{a}{\ Positions and uncertainties were calculated as
  described in Section \ref{sec:cl_properties}.}
\tablenotetext{b}{\ $Y_{0\farcm75}$ is the integrated Comptonization
  within a $0\farcm75$ radius, as described in section
  \ref{sec:cl_properties}.  Statistical errors are reported in this
  table to reflect the S/N.  Calibration uncertainties
  add an additional systematic error of $\approx10\%$.}
\tablenotetext{c}{\ Optical coordinates are given for the BCG from each cluster.}
\tablenotetext{d}{\ $z$ refers to red-sequence redshifts.  Uncertainties are $\sigma_z / (1+z) \approx 2$--$3\%$ for all clusters except \sptwhopper, whose uncertainty is $z$ is $\sigma_z / (1+z) \approx 5\%$}
\tablenotetext{e}{\ X-ray luminosities are reported in the rest frame 0.5-2.0 keV energy band.}
\end{deluxetable*}